\documentclass{appolb}
\usepackage{epsfig}

\begin{document}
\title{Flow characteristics and strangeness production in the framework of highly-anisotropic and strongly-dissipative hydrodynamics%
\thanks{Presented at \textit{Strangeness in Quark Matter 2011}, Sept. 18--24, Cracow, Poland.}%
}
\author{Rados{\l }aw Ryblewski
  \address{The H. Niewodnicza\'nski Institute of Nuclear Physics,\\
    Polish Academy of Sciences,\\
    ul. Radzikowskiego 152, PL-31342 Krak\'ow, Poland}
}
\maketitle
\begin{abstract}
The recently formulated model of highly-anisotropic and strongly dissipative hydrodynamics is used in 3+1 dimensions to describe flow characteristics and strangeness production in Au+Au collisions at the highest RHIC beam energy. Our results show very weak dependence on the initial momentum anisotropy, provided the anisotropic phase lasts no longer than 1 fm/c.  
\end{abstract}
  
\section{Introduction}
Soft-hadronic data collected in the ultra-relativistic heavy-ion experiments may be well described in the framework of the standard perfect-fluid hydrodynamics \cite{Florkowski:2010zz} or by dissipative hydrodynamics with small viscosity \cite{Chaudhuri:2006jd,Dusling:2007gi,Luzum:2008cw,Song:2007fn,Bozek:2009dw,Schenke:2010rr}. However, the success of those approaches relies strongly on the assumption that the produced system reaches state of local thermal equilibrium within a fraction of a fermi\footnote{We use the natural system of units where $\hbar=c=1$.}. 

Most of the microscopic models of the early stages fail to explain such short thermalization times \cite{Florkowski:2010zz}. This difficulty is known as the \textit{early thermalization puzzle}. One of its common solutions is the concept of a strongly coupled quark-gluon plasma \cite{Shuryak:2004cy}. In addition, many microscopic approaches assume that the produced system exhibits initially large anisotropies in the momentum space, for example, see \cite{Bjoraker:2000cf}. 

Recently, several models have been developed \cite{Broniowski:2008qk,Ryblewski:2010tn}, which include the early highly-anisotropic phase of the collisions. This has been achieved introducing a pre-equilibrium stage connected with a subsequent perfect-fluid description. Combination of different approaches in a single framework seems, however, not completely satisfactory. The need for a concise model which can describe different stages of heavy-ion collisions in the uniform way triggered our development of the highly-Anisotropic and strongly-Dissipative HYDROdynamics (ADHYDRO) \cite{Florkowski:2010cf,Ryblewski:2010bs,Ryblewski:2010ch,Ryblewski:2011aq}, see also \cite{Martinez:2010sc,Martinez:2010sd}

In this paper we present our results obtained within the ADHYDRO framework coupled to THERMINATOR \cite{Kisiel:2005hn,Chojnacki:2011hb}. For the first time we use ADHYDRO in 3+1 dimensions. We discuss flow characteristics of the emitted hadrons and strangeness production depending on the initial momentum anisotropy. 

\section{Formulation of the model}

In the ADHYDRO model the evolution of the system is described by the following equations \cite{Florkowski:2010cf}
\begin{eqnarray}
\partial_\mu T^{\mu \nu} &=& 0, \label{enmomcon} \\
\partial_\mu \sigma^{\mu} &=& \Sigma, \label{engrow}
\end{eqnarray}
which express the energy-momentum conservation and the entropy production laws, respectively.
The energy-momentum tensor $T^{\mu \nu}$ in Eq.~(\ref{enmomcon}) has the form
\begin{eqnarray}
T^{\mu \nu} = \left( \varepsilon  + P_{\perp}\right) U^{\mu}U^{\nu} - P_{\perp} \, g^{\mu\nu} - (P_{\perp} - P_{\parallel}) V^{\mu}V^{\nu}, 
\label{Taniso}
\end{eqnarray}
which allows for the asymmetry between  the longitudinal, $P_{\parallel}$,  and transverse, $P_{\perp}$, pressures. In the limit where the system becomes isotropic,  $P_{\parallel}=P_{\perp}=P$, the formula (\ref{Taniso}) reproduces the energy-momentum tensor of the perfect fluid. Similarly, the entropy production law (\ref{engrow}) is reduced to the entropy conservation law, if we assume that the entropy source, $\Sigma$, vanishes. The four-vector $U^{\mu}$ defines the four-velocity of the fluid and $V^{\mu}$ is the four-vector defining the beam axis. In the general case, $U^{\mu}$ and $V^{\mu}$ may be parametrized in the following way
\begin{eqnarray}
U^\mu &=& (u_0 \cosh \vartheta, u_x, u_y, u_0 \sinh \vartheta), \label{U3+1} \\
V^\mu &=& (	 \sinh \vartheta, 0, 0,  \cosh \vartheta), \label{V3+1}
\end{eqnarray}
where $u_x$ and $u_y$ are the transverse components of the four-velocity velocity field ($u_\perp = \sqrt{u_x^2 + u_y^2}$ and $u_0 = \sqrt{1+u_\perp^2}$), $\vartheta$ is the longitudinal fluid rapidity. The parametrizations (\ref{U3+1}) and (\ref{V3+1}) satisfy simple normalization conditions, i.e., $U^2 = 1$, $V^2 = -1$, $U \cdot V = 0$.
The entropy flux $\sigma^{\mu}$ in (\ref{engrow}) is defined by the formula
\begin{equation}
\sigma^{\mu} = \sigma \, U^\mu,
\label{Saniso}
\end{equation}
where $\sigma$ is the non-equilibrium entropy density.

One can show \cite{Florkowski:2010cf} that instead of $P_{\parallel}$ and $P_{\perp}$ it is more convenient to use the entropy density $\sigma$ and the anisotropy parameter $x$ as two independent variables (to a good approximation we have $P_{\parallel}/P_{\perp} = x^{-3/4}$). Similarly to standard hydrodynamics with vanishing baryon chemical potential, the energy density $\varepsilon$ introduced in Eq.~(\ref{Taniso}), the entropy density  $\sigma$, and the anisotropy parameter $x$ are related through the \textit{generalized} equation of state $\varepsilon=\varepsilon(\sigma, x)$. In our model we use the following ansatz \cite{Ryblewski:2010bs}
\begin{eqnarray}
\varepsilon (x,\sigma)&=&  \varepsilon_{\rm qgp}(\sigma) r(x), \label{epsilon2b}  \\ \nonumber 
P_\perp (x,\sigma)&=&  P_{\rm qgp}(\sigma) \left[r(x) + 3 x r^\prime(x) \right], \label{PT2b}   \\ \nonumber 
P_\parallel (x,\sigma)&=&  P_{\rm qgp}(\sigma) \left[r(x) - 6 x r^\prime(x) \right]. \label{PL2b} 
\end{eqnarray}
where $\varepsilon_{\rm qgp}$ and  $P_{\rm qgp}$  define the realistic equation of state constructed in Ref. \cite{Chojnacki:2007jc}. The function $r(x)$ is the pressure relaxation function characterizing the properties of the fluid which exhibits the anisotropy $x$.  Here we use the form introduced in \cite{Florkowski:2010cf}
\begin{equation}
r(x) = \frac{ x^{-\frac{1}{3}}}{2} \left[ 1 + \frac{x \arctan\sqrt{x-1}}{\sqrt{x-1}}\right].
\label{RB}
\end{equation}
In the isotropic case $x = 1$, $r(1)=1$, $r^\prime(1)=0$, and Eq.~(\ref{epsilon2b}) is reduced to the equation of state used in \cite{Chojnacki:2007jc}.

The function $\Sigma$ which appears on the right-hand-side of Eq.~(\ref{engrow}) defines the entropy production due to microscopic processes taking place in the system. Exactly these processes lead to thermalization of the system. We use the form proposed in \cite{Florkowski:2010cf}
\begin{equation}
\Sigma(\sigma,x) = \frac{(1-\sqrt{x})^{2}}{\sqrt{x}}\frac{\sigma}{\tau_{\rm eq}},
\label{entropys}
\end{equation}
where the time-scale parameter $\tau_{\rm eq}=0.25$ fm controls the rate of equilibration\footnote{Using this value we find that the system equilibrates within about 1 fm.}. In the limit of small anisotropy Eq.~(\ref{entropys}) is consistent with the quadratic form of the entropy production in the Israel-Stewart theory. Far from equilibrium, hints for the form of $\Sigma$ are lacking, although we may expect some suggestions from the AdS/CFT correspondence \cite{Heller:2011ju}. Thus, for large anisotropies the formula (\ref{entropys}) should be treated as an assumption defining the dynamics of the system. 
%
\section{Initial conditions and freeze-out}
\label{sec_ini}
In the general 3+1 case we have to solve Eqs. (\ref{enmomcon}) and (\ref{engrow}) for five unknown functions $\sigma$, $x$, $u_x$, $u_y$, and $\vartheta$, which depend on the space-time coordinates: $\tau,{\bf x}_\perp$, and $\eta$ ($\tau$ is the proper time and $\eta$ is the space-time rapidity). Since the system's evolution is treated hydrodynamically from the very early stages where the anisotropies are expected to be very large, we fix the initial starting time for ADHYDRO to $\tau_{\rm 0} =0.25$ fm. Similarly to other hydrodynamic calculations, we assume that there is no initial transverse flow, $u_x(\tau_{\rm 0},{\bf x}_\perp,\eta) = u_y(\tau_0,{\bf x}_\perp,\eta) = 0$. For the initial longitudinal rapidity of the fluid we assume the Bjorken scaling $\vartheta(\tau_0,{\bf x}_\perp,\eta) = \eta$. We check three scenarios: i) the initial source is strongly oblate in the momentum space, $x(\tau_0,{\bf x}_\perp,\eta) =100$, which corresponds to a transversally thermalized source, ii) the source is prolate in momentum space, $x(\tau_0,{\bf x}_\perp,\eta) = 0.032$ (this case is analogous to i) because $r(100) = r(0.032)$), and iii) the initial source is isotropic, $x(\tau_0,{\bf x}_\perp,\eta) =1$, which gives the closest description to the standard hydrodynamic case. The initial entropy density profile is given by the formula
\begin{equation}
 \sigma(\tau_0,{\bf x}_\perp,\eta) = 
\varepsilon_{\rm gqp}^{-1} 
\left[ \frac{\varepsilon_{\rm i} \, \tilde{\rho}(b,{\bf x}_\perp,\eta)}{r(x(\tau_0,{\bf x}_\perp,\eta))} \right], \quad \quad
 \tilde{\rho}(b,{\bf x}_\perp,\eta) = \frac{\rho(b,{\bf x}_\perp,\eta)}{\rho(0,0,0)},
\label{densinit}
\end{equation}
where $\tilde{\rho}$ is the normalized initial density of sources. The density profile $\rho$ is given as the tilted source worked out in Ref. \cite{Bozek:2010bi} and $\varepsilon_{\rm gqp}^{-1}$ is the inverse of the function $\varepsilon_{\rm gqp}(\sigma)$. The initial energy density $\varepsilon_{\rm i}= 107.5$ $ \mathrm{GeV/fm^3}$ is the same for all three analyzed cases. 

The evolution is determined by the hydrodynamic equations until the entropy density drops to $\sigma_{\rm f} = 1.79$ fm$^{-3}$, which for $x=1$ corresponds to the temperature $T_{\rm f} = 150$ MeV. According to the single-freeze-out scenario, at this moment the abundances and momenta of particles are expected to freeze-out and particles freely stream to detectors. The processes of particle production and decays of unstable resonances are described by using  {\tt THERMINATOR 2} \cite{Chojnacki:2011hb}, which applies the Cooper-Frye formalism to generate hadrons on the freeze-out hypersurface extracted from ADHYDRO.
%
\section{Results and conclusions}

The {\tt THERMINATOR 2} package \cite{Chojnacki:2011hb} allows us the calculation of several one- and two-particle observables (such as the particle $p_{\rm T}$ spectra, the directed and elliptic flow coefficients, the HBT radii, etc.). Due to limited space, in this paper we show only the hyperon spectra and the directed flow coefficient $v_1$. More results will be presented and discussed in a separate publication.

\begin{figure}[t]
\begin{center}
\includegraphics[angle=0,width=0.49\textwidth]{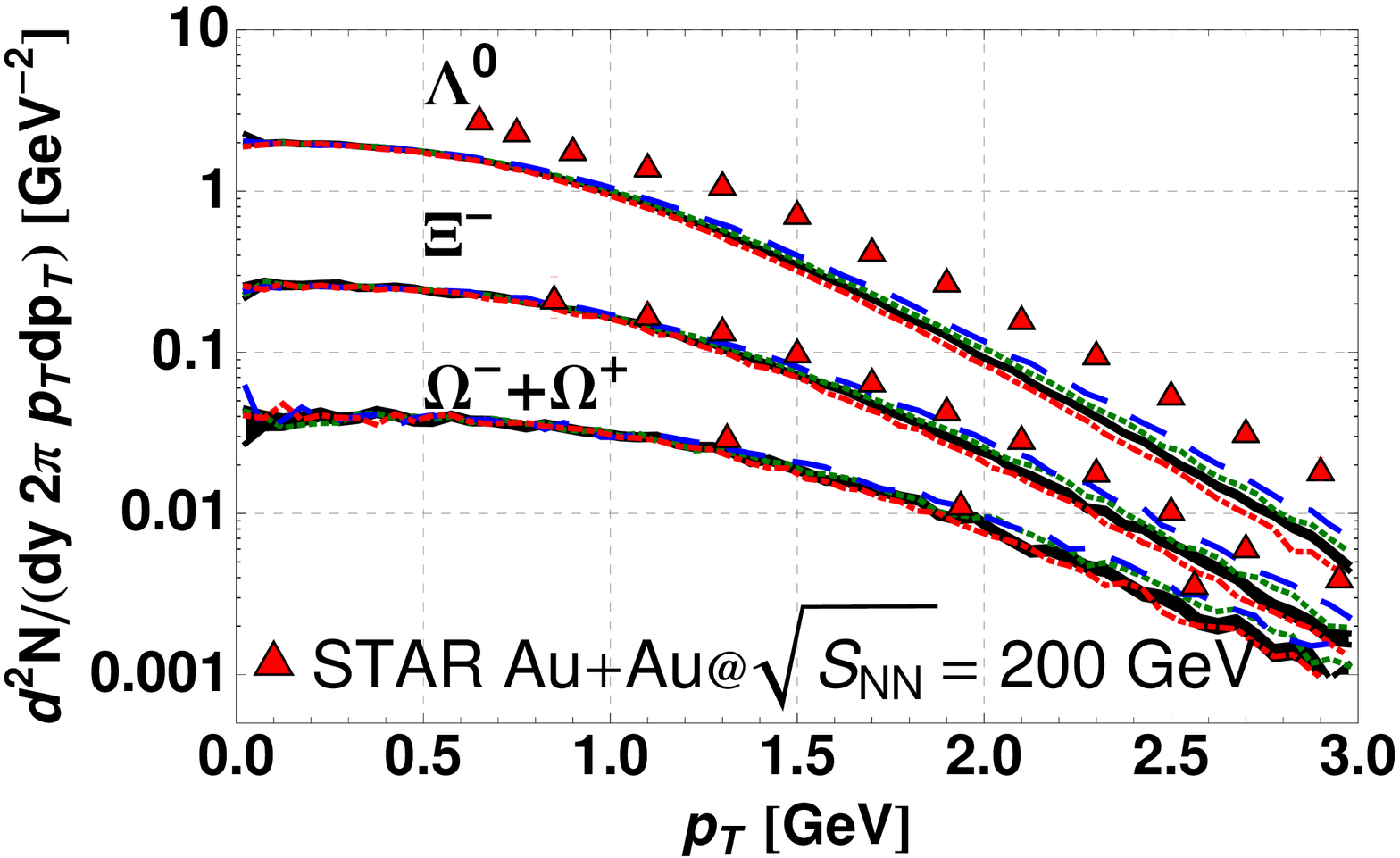}
\includegraphics[angle=0,width=0.49\textwidth]{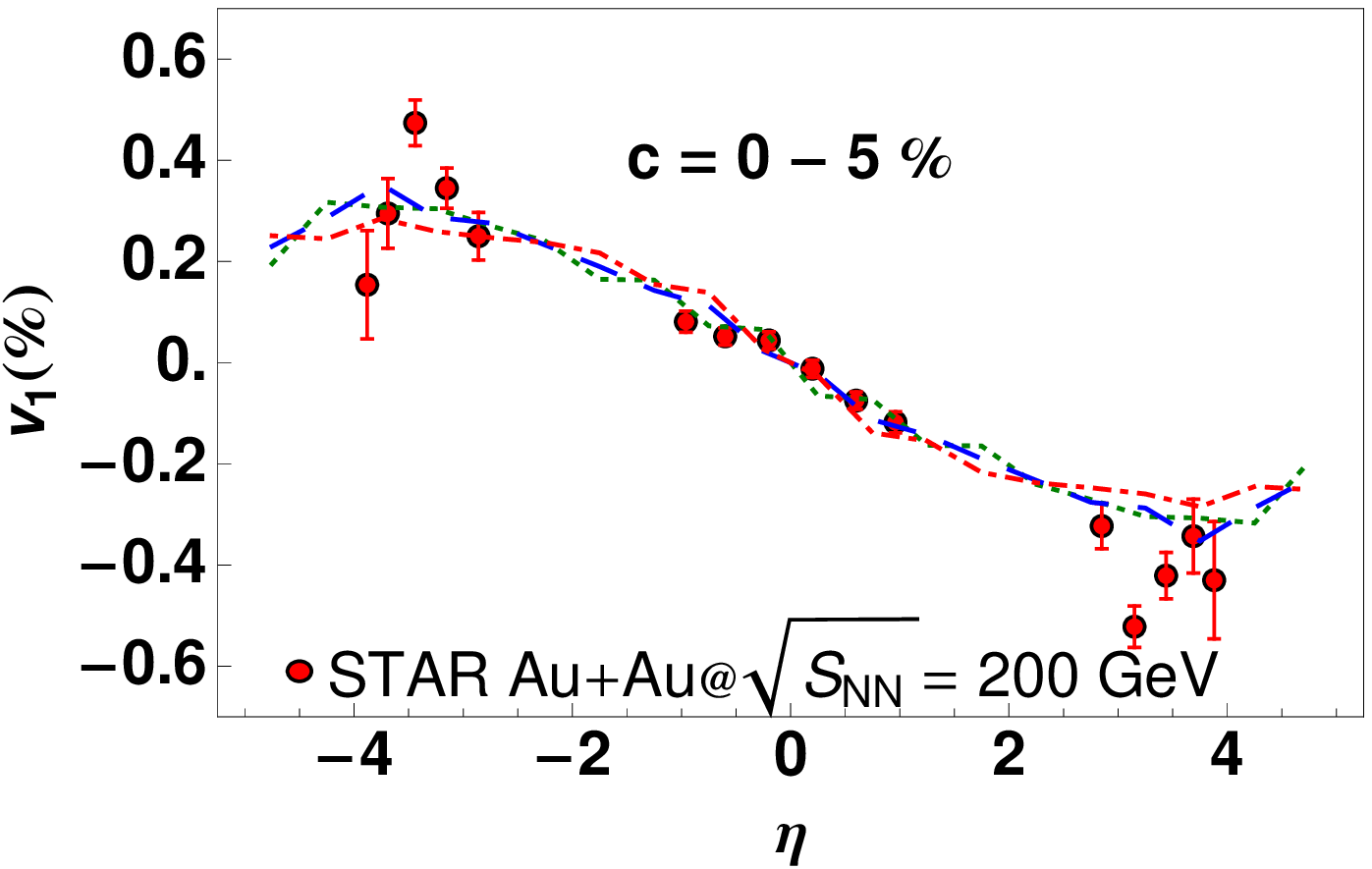}
\end{center}
\caption{\small Transverse-momentum spectra of hiperons (left part) compared to the experimental data from STAR \cite{Adams:2006ke} and directed flow coefficient (right part) compared to the STAR data \cite{Abelev:2008jga}. Theoretical lines are obtained from ADHYDRO for Au+Au collisions at $\sqrt{s_{\rm NN}}=200$ GeV and the centrality class $c=0-5$\%. The results are shown for three cases: oblate source (blue dashed line), prolate source (red dashed-dotted line) and perfect fluid (green dotted line) as discussed in Sect.~\ref{sec_ini}. }
\label{fig:spectra0005}
\end{figure}

The scaling (\ref{densinit}) helps us to keep the final particle multiplicities approximately the same in the three considered cases. In particular, this can be concluded from the left-hand-side of Fig.~\ref{fig:spectra0005} where we show the transverse momentum spectra of hyperons. We observe that the stronger transverse flow in the case i) results in a bit harder spectra as compared to the case ii). Despite this fact, the spectra do not differ significantly. We observe that the model spectra of $\Xi$'s and $\Omega$'s agree well with the data, while the normalization of $\Lambda$'s is too small.

The authors of Ref.~\cite{Bozek:2010aj} have shown that the $v_1$ coefficient may be treated as a probe for measuring the thermalization time, since it is very sensitive to the early difference of pressures. On the right-hand-side of Fig.~\ref{fig:spectra0005} we present our results for the directed flow coefficient $v_1$. We observe small sensitivity to large initial anisotropy in the midrapidity region provided the anisotropic stage lasts not longer than 1 fm. Our results are not so much restrictive as the results presented in \cite{Bozek:2010aj}, since we do not fix the final multiplicities but allow them to vary within the experimental errors. 

More results obtained in the ADHYDRO model in 3+1 dimensions will be presented in a separate publication. We note that our previous results obtained in the 2+1 version \cite{Ryblewski:2011aq} (a boost-invariant version) show similar, weak dependence of other physical observables on the initial pressure asymmetry. In conclusion, we state that we have found further evidence that there is a place for a highly-anisotropic phase at the early stages of heavy-ion collisions, provided the system reaches local thermal equilibrium before 1 fm.


\section{Acknowledgments}
This work was supported in part by the MNiSW grant No. N N202 288638.

\begin{thebibliography}{20}

\bibitem{Florkowski:2010zz}
W. Florkowski, {\it Phenomenology of Ultra-Relativistic Heavy-Ion Collisions}, World Scientific, Singapore, 2010.

\bibitem{Chaudhuri:2006jd}
A.~K.~Chaudhuri, {\it Phys. Rev.} {\bf C74}, 044904 (2006).

\bibitem{Dusling:2007gi}
K.~Dusling, D.~Teaney, {\it Phys. Rev.} {\bf C77}, 034905 (2008).

\bibitem{Luzum:2008cw}
M.~Luzum, P. Romatschke, {\it Phys. Rev.} {\bf C78}, 034915 (2008).

\bibitem{Song:2007fn}
H.~Song, U.~Heinz, {\it Phys. Lett.} {\bf B658}, 279 (2008)

\bibitem{Bozek:2009dw}
P.~Bozek, {\it Phys. Rev.} {\bf C81}, 034909 (2010).

\bibitem{Schenke:2010rr}
B.~Schenke, S.~Jeon, C.~Gale, {\it Phys. Rev. Lett.} {\bf 106}, 042301 (2011).

\bibitem{Shuryak:2004cy}
E.~V.~Shuryak, {\it Nucl. Phys.} {\bf A750}, 64 (2005).

\bibitem{Bjoraker:2000cf}
J.~Bjoraker, R.~Venugopalan, {\it Phys. Rev.} {\bf C63}, 024609 (2001).

\bibitem{Broniowski:2008qk}
W.~Broniowski, W.~Florkowski, M.~Chojnacki, A.~Kisiel, \textit{Phys. Rev.} {\bf C80}  034902 (2009).

\bibitem{Ryblewski:2010tn}
R.~Ryblewski, W.~Florkowski, \textit{Phys. Rev.} {\bf C82}  024903 (2010).

\bibitem{Florkowski:2010cf}
W.~Florkowski, R.~Ryblewski, {\it Phys. Rev.} {\bf C83}, 034907 (2011).

\bibitem{Ryblewski:2010bs}
R.~Ryblewski, W.~Florkowski, {\it J. Phys.} {\bf G38}, 015104 (2011).

\bibitem{Ryblewski:2010ch}
R.~Ryblewski, W.~Florkowski, {\it Acta Phys. Pol.} {\bf B42}  115 (2011).

\bibitem{Ryblewski:2011aq}
R.~Ryblewski, W.~Florkowski, arXiv:1103.1260.

\bibitem{Martinez:2010sc}
M.~Martinez, M.Strickland, {\it Nucl. Phys.} {\bf A848}, 183 (2010).

\bibitem{Martinez:2010sd}
M.~Martinez, M.Strickland, {\it Nucl. Phys.} {\bf A856}, 68 (2011).

\bibitem{Kisiel:2005hn}
A.~Kisiel et al., {\it Comput.Phys.Commun.}
{\bf 174}, 669 (2006).

\bibitem{Chojnacki:2011hb}
M.~Chojnacki et al., arXiv:1102.0273.

\bibitem{Chojnacki:2007jc}
M.~Chojnacki, W.~Florkowski, \textit{Acta Phys. Pol.} {\bf B38}  3249 (2007).

\bibitem{Heller:2011ju}
M. ~Heller, R.~Janik, P.~Witaszczyk, arXive:1103.3452.



%
\bibitem{Bozek:2010bi}
P.~Bozek, I.~Wyskiel, {\it Phys. Rev.} {\bf C81}, 054902 (2010).
%

%
\bibitem{Adams:2006ke}
Adams et al. (STAR Collaboration),\textit{Phys. Rev. Lett.} \textbf{98}, 62301 (2007). 
%
\bibitem{Abelev:2008jga}
B.I.~Abelev et al. (STAR Collaboration), \textit{Phys. Rev. Lett.} \textbf{101}, 252301 (2008). 
%

%
\bibitem{Bozek:2010aj}
P.~Bozek, I.~Wyskiel-Piekarska, {\it Phys. Rev.} {\bf C83}, 024910 (2011).



\end{thebibliography}
\end{document}